\documentclass[a4paper,11pt]{article}
\usepackage{amsmath,amssymb,amsfonts,graphicx}

\newcommand{\phibar}{\bar{\phi}}
\newcommand{\Tr}{\mathop{Tr}\nolimits}
\begin{document} 

\begin{flushright} 
DFTT44/2009\\
WIS/09/09-JUNE-DPP\\
RIKEN-TH-161

\end{flushright} 

\vspace{0.1cm}

\begin{center}
    \noindent{\LARGE{
Lattice study of\\ two-dimensional ${\cal N}=(2,2)$ super Yang-Mills\\ at large-$N$
}}\\
    
    \vspace{2cm} 
    
     \noindent{
 \def\thefootnote{\fnsymbol{footnote}}
       Masanori Hanada$^{ab}$\footnote{e-mail:masanori.hanada@weizmann.ac.il} and 
       Issaku Kanamori$^{bc}$\footnote{e-mail:kanamori@to.infn.it}}\\
    \setcounter{footnote}{0}
    \vspace{1cm}	
    $^{a}${\it Department of Particle Physics,  
Weizmann Institute of Science 

Rehovot 76100, Israel} \\
	 $^b$ {\it Theoretical Physics Laboratory,
RIKEN Nishina Center

 Wako, Saitama 351-0198, Japan}\\
 
		$^c$
{\it Dipartimento di Fisica Teorica, Universit\`a di Torino,
Via Giuria 1, 10125 Torino, Italy}\\ 

    \vskip 3cm
  \end{center}

\begin{abstract}
We study two-dimensional ${\cal N}=(2,2)$ $SU(N)$ super Yang-Mills theory 
on Euclidean two-torus using Sugino's lattice regularization. 
We perform the Monte-Carlo simulation for $N=2,3,4,5$ and then 
extrapolate the result to $N=\infty$. With the periodic boundary conditions 
for the fermions along both circles, we establish the existence of a bound state 
in which scalar fields clump around the origin, in spite of the existence of 
a classical flat direction. In this phase the global $({\mathbb Z}_N)^2$ symmetry 
turns out to be broken. We provide a simple explanation for this fact 
and discuss its physical implications. 

\end{abstract} 

\newpage
\section{Introduction}
Large-$N$ supersymmetric Yang-Mills theories (SYM) 
are promising candidates for nonperturbative formulations of 
the superstring theories~\cite{BFSS96,IKKT96,DVV97,Maldacena97,IMSY98}.  
Recent developments of discretization technique enable us to study 
nonperturbative aspects of these theories. (For a recent review, see e.g.~%
\cite{Giedt07,CKU09}.) 
Especially, one-dimensional 
maximally supersymmetric gauge theory has been studied extensively~%
\cite{HNT07,AHNT07,CW07,CW08,HMNT08} 
and the gauge/gravity duality~\cite{Maldacena97,IMSY98} has been confirmed 
very precisely, including the stringy $\alpha'$ corrections~\cite{HMNT08}. 
By assuming that the gauge/gravity duality holds, then the Monte-Carlo 
simulation of the gauge theory provides a new and powerful tool to study 
the physics of black holes. 

Next simplest model to study is two-dimensional theory, 
of which some lattice regularizations are known. 
Maximally supersymmetric 2d SYM  is expected to have dual 
D1-brane description in type IIB superstring theory~\cite{IMSY98}\footnote{
Of course this model has another interesting interpretation as the 
``matrix string theory''~\cite{DVV97}. 
}.  
By compactifying the spatial direction, one obtains D1-branes 
winding on the compactified direction. By taking T-dual, one obtains a system of 
D0-branes in compact space, which can have several phases -- 
if D0-branes are smeared along the compactified direction it is a
\emph{black string}, 
and if they are localized in a small region it is a \emph{black hole}. 
The transition between these phases (the Gregory-Laflamme 
transition~\cite{GL93}) corresponds 
to the breakdown of the global ${\mathbb Z}_N$ symmetry  
in the gauge theory~\cite{GL_dual gauge theory}.  
By studying the Gregory-Laflamme transition in supergravity and 
then by using the gauge/gravity duality to translate the supergravity to the gauge theory, 
one can study the phase structure of the gauge theory at strong coupling~\cite{AMMPRW05}. 
Using the duality in the opposite direction, we can study the detail of 
the stringy correction to the Gregory-Laflamme transition 
with the Monte-Carlo simulation of the gauge theory.  
At present, it is difficult to study the maximally supersymmetric 2d SYM by using the Monte-Carlo 
simulation. However, four-supercharge system has been studied extensively by using 
the formulations free from fine tuning. 

In this paper, we study two-dimensional ${\cal N}=(2,2)$ $SU(N)$ SYM on two-torus 
using Sugino's lattice regularization~\cite{Sugino04}\footnote{
For other regularizations of this theory, see~\cite{2d other formulations}. 
}.  
In this model, the restoration of supersymmetry without fine-tuning has been tested 
extensively \cite{KS08sono2}\footnote{
See also \cite{Suzuki07,KSS07,KS08,Kanamori09}.   
For simulations using other formulations, see e.g. \cite{Catterall08}. 
}.  
We study $N=2,3,4, 5$ and extrapolate to the planar limit $N=\infty$. 
The action in the continuum 
is obtained from 4d ${\cal N}=1$ SYM through the dimensional reduction,  
and is given by 
\begin{equation}
S
=
N\int_0^{L_x} dx\int_0^{L_y}dy
\ \Tr
\left\{
\frac{1}{4}F_{\mu\nu}^2
+
\frac{1}{2}(D_\mu X_i)^2
-
\frac{1}{4}[X_i,X_j]^2
-
\frac{i}{2}\bar{\psi}\Gamma^\mu D_\mu\psi
-
\frac{1}{2}\bar{\psi}\Gamma^i [X_i,\psi]
\right\}, 
\end{equation}
where $\mu$ and $\nu$ run $x$ and $y$, $i$ and $j$ run $1$ and $2$, 
and $\Gamma^I=(\Gamma^\mu,\Gamma^i)$ are gamma matrices in four dimensions.   
$X_i$ are $N\times N$ hermitian matrices, $\psi_\alpha$ are $N\times N$ 
fermionic matrices with a Majorana index $\alpha$ 
and the covariant derivative is given by $D_\mu=\partial_\mu-i[A_\mu,\ \cdot\ ]$. 
The only parameters of the model are 
the size of circles $L_x$ and $L_y$. (Note that the coupling constant 
can be absorbed by redefining the fields and coordinates. 
In other words, the 't~Hooft coupling $\lambda$ which has mass dimension
$2$ can be taken to be $\lambda=1$. 
Then the strong coupling corresponds to the large volume.)
There are several motivations to study this system. Firstly, 
it is the simplest SYM in two dimensions which can be studied 
nonperturbatively by lattice simulation. 
Especially, notorious ``sign problem'' is absent.  
Secondly, we can expect that it is qualitatively similar 
to maximally supersymmetric (${\cal N}=(8,8)$) SYM, 
which is conjectured to be dual to type II superstring. 
Thirdly, its bosonic cousin is studied in~\cite{NN03} 
and it is interesting to compare the phase structure. 
In the bosonic model, the ${\mathbb Z}_N$ symmetry is broken below the critical volume. 
In the supersymmetric model it is expected to be broken in any finite volume~\cite{HMNR09}. 
The argument in~\cite{HMNR09} is valid only large and small volume region and 
it is desirable to check the breakdown of ${\mathbb Z}_N$ at intermediate volume.  

An obstacle for the simulation is the existence of the flat direction, 
along which two scalar fields $X_1$ and $X_2$ commute. 
In contrary to a theory on ${\mathbb R}^{1,3}$, there is no superselection of the moduli 
parameter in this case. That is, eigenvalues of scalars are determined dynamically. 
Therefore, some mechanism which restrict eigenvalues to a finite region 
is necessary in order for the stable simulation. 
One possible way is to introduce an IR regulator 
and gradually remove it. 
In~\cite{KS08sono2,KS08,Kanamori09} this method has been applied 
to the $SU(2)$ theory at finite temperature. In those works, 
a mass term of scalars has been introduced and physical quantities 
are evaluated by an extrapolation to the massless limit or evaluated
with small scalar mass. 
In this case, as we will see,  
in fact the scalar eigenvalues spread as IR regulator is removed. 
In string terminology, this phase can be understood as a gas of freely propagating D-branes. 

At large-$N$, there is a more interesting phase, namely a bound state  
in which eigenvalues clump to a small region~\cite{HMNR09}. 
It is metastable at finite-$N$ and becomes stable at large-$N$. 
This bound state is a cousin  
of the one in one-dimensional system, which has been found in~\cite{AHNT07}, 
and corresponds to the black brane background in type II superstring. 
(The black brane solution in supergravity corresponds to the bound state 
of the D-branes. Because the scalar fields represent the collective 
coordinates of the D-branes, the bunch of the D-branes is nothing but  
the bound state of scalar eigenvalues.) 
In this paper we study this phase. 
Because the very existence of this bound state is nontrivial from 
a purely field-theoretical point of view, we provide numerical evidence. 
As we will see, with periodic-periodic (P-P) boundary conditions for fermions, 
we construct the bound state explicitly.  
In this bound state, there are two possible phases, namely the $({\mathbb Z}_N)^2$ 
broken and unbroken phases. 
(The model has global $({\mathbb Z}_N)^2$ symmetry, which shifts the 
complex phases of 
Wilson loops winding on circles. If Wilson loops are non-zero, 
then the $({\mathbb Z}_N)^2$ symmetry is spontaneously broken.) 
We study the phase structure under the variation of the periods. 
Because of the limitation of the resources, we study only the case of 
$L_x=L_y$.  
We confirm that, in the bound state, 
the $({\mathbb Z}_N)^2$ symmetry is broken as discussed in~\cite{HMNR09}. 
With antiperiodic-periodic (A-P) boundary conditions, 
we need rather large $N$ to find such an bound state and we could not 
construct it numerically. Hence we cannot discuss the thermal properties of 
the black brane in this paper. 
However we can expect that we can study the finite temperature 
system in near future, by using a faster computer. 

This paper is organized as follows. 
In~\S~\ref{sec:phase diagram} we review the conjectured phase diagram and 
explain its physical implications. 
In~\S~\ref{sec:simulation} we show the numerical results. 
We give the conclusion and discussion in~\S~\ref{sec:conclusion}.
The simulation details are explained in the appendices.

\section{Conjectured phase structure}\label{sec:phase diagram}
In this section we review the expectations on the phase structure.
In short, the ${\mathbb Z}_N$ symmetry is broken in the bound state. 
The phases are summarized in Table~\ref{table:phase}. 

\begin{table}
 \begin{tabular}{l|c|c|c}
  \multicolumn{1}{c|}{Model} & large volume & finite volume & small volume \\
  \hline
  4d $\mathcal{N}=1$, P-P     
     & $(\mathbb{Z}_N)^4$ and (broken) 
     & $(\mathbb{Z}_N)^4$ and (broken) 
     & $(\mathbb{Z}_N)^4$ and (broken) \\
  2d $\mathcal{N}=(2,2)$, P-P 
     & $(\mathbb{Z}_N)^2$ 
     & (broken) 
     & (broken)\\
  2d $\mathcal{N}=(8,8)$, A-P
     & $\mathbb{Z}_N$
     & (phase transition(s))
     & (broken) \\
 \end{tabular}
\caption{Conjecture for phases of various super Yang-Mills.  
 In this paper, we confirm
 the 2d $\mathcal{N}=(2,2)$ case numerically. }\label{table:phase}
\end{table} 
\subsection{4d ${\cal N}=1$ SYM on Euclidean four-torus}
Let us start by considering the 4d ${\cal N}=1$ SYM on four-torus $T^4$. 
For all circles we impose periodic boundary conditions for both bosons and fermions. 
Let $L_\mu\ (\mu=1,\cdots,4)$ be the periods along four directions and  
$W_\mu$ be the Wilson loops winding on each directions, 
\begin{eqnarray}
W_\mu = P\exp\left(i\int_0^{L_\mu} dx^\mu A_\mu\right), 
\end{eqnarray}
where the contraction over $\mu$ is not taken in the right hand side. 
Under the global $({\mathbb Z}_N)^4$ transformation, $W_\mu$ are multiplied 
by phase factors, 
\begin{eqnarray}
W_\mu \to e^{2\pi in_\mu/N}W_\mu
\qquad
(n_\mu\in {\mathbb Z}). \label{eq:phase-factor}
\end{eqnarray}
Therefore the $({\mathbb Z}_N)^4$ is broken if $\Tr W_\mu$ have nonzero expectation values. 
In 4d ${\cal N}=1$, there are (at least) two phases -- $({\mathbb Z}_N)^4$ broken and unbroken. 
(Note that the ${\mathbb Z}_N$ symmetry can be broken only in the large-$N$ limit.)

The existence of the $({\mathbb Z}_N)^4$ unbroken phase can be shown 
as follows~\cite{KUY07}. Consider the situation in which one of the periods $L_1$ is 
much smaller than other three periods $L_2,L_3$ and $L_4$. 
Then essentially we obtain SYM on ${\mathbb R}^3\times S^1$. 
If we take $L_1\ll \Lambda_{{\rm QCD}}^{-1}\ll L_2,L_3,L_4$, 
the effective potential of the Wilson line phase $\ln W_1$ 
can be calculated perturbatively and it gives exactly zero.    
By taking monopole and instanton effects~\cite{DHKM99} into account, 
it turns out that the eigenvalue of the Wilson line 
(the Wilson line phases) repel each other once they spread uniformly. 
Therefore, ${\mathbb Z}_N$-unbroken configuration is stable. 
This calculation itself is done in a specific limit, but 
due to the supersymmetry the same result should hold at any volume~\cite{KUY07}. 

The $({\mathbb Z}_N)^4$ broken phase can be found at small volume 
limit~\cite{HMNR09}. 
Suppose that $({\mathbb Z}_N)^4$ is broken, or equivalently, 
the Wilson line phases clump in a small region around the origin. 
If the size of the phase distribution is small enough, 
then the system can be approximated by its zero-dimensional
reduction\footnote{
In the dimensional reduction, KK modes are neglected while the effective
mass term coming from the commutator term $\Tr [A_\mu,A_\nu]^2$ in the field 
strength term is kept. In order for this approximation to make sense, 
the effective mass must be sufficiently smaller than the KK mass, that
is, eigenvalues of $A_\mu$ must be small enough. Then the Wilson loop is
close to 1 and hence the 
${\mathbb Z}_N$ symmetry is broken. 

}. 
This model has been studied extensively. 
Especially, it is known to have a bound state of eigenvalues even though 
it has a flat direction classically, and 
the size of the eigenvalue 
distribution is known~\cite{KS99,AABHN00,Austing:2001pk}. 
It is small enough 
so that the assumption that the system is described by the zero-dimensional 
reduction is correct. Therefore, a $({\mathbb Z}_N)^4$ broken phase exists 
at small volume. We expect this phase persists at any finite volume 
due to the supersymmetry. 
\subsection{2d ${\cal N}=(2,2)$ SYM on two-torus}\label{subsec:2dSYM}
2d ${\cal N}=(2,2)$ SYM is obtained from 4d ${\cal N}=1$ through 
the dimensional reduction. 
When we take two circles in 4d theory to be small, in order 
for the dimensional reduction to work the ${\mathbb Z}_N$ symmetries  
along these circles must be broken. In such a phase Wilson line phases 
clump, so it corresponds to the bound state in 2d theory. 
Therefore, it is natural to assume that 
the bound state in 2d theory corresponds 
to $({\mathbb Z}_N)^4$ broken phase in 4d theory. 
If this is the case, the $({\mathbb Z}_N)^2$ symmetry of the 2d theory should be broken. 

At large volume, $({\mathbb Z}_N)^2$ symmetry should be restored. 
This is because $({\mathbb Z}_N)^2$ is essentially continuous 
$U(1)^2$ symmetry at large-$N$ and hence at noncompact two-dimensional 
space it is unbroken according to the Coleman's theorem. 

In \S~\ref{sec:simulation}, 
we will confirm the above statement by numerical simulation. 
The discussion above does not exclude an existence of 
a ${\mathbb Z}_N$ unbroken phase; 
see a remark in \S~\ref{sec:bound state}. 
\subsection{2d ${\cal N}=(8,8)$ SYM : ${\mathbb Z}_N$ broken phase as a black hole}

For 2d ${\cal N}=(8,8)$ SYM, there is a dual gravity interpretation. 
Here we consider the correspondence between thermal SYM and 
black branes in type II supergravity.  
First, let us briefly describe the SYM in terms of D-branes. 
We impose the antiperiodic boundary condition. for temporal direction 
(we take $y$ to be temporal direction). 
In this case, the bound state (i.e. a state in which 
scalar eigenvalues clump around the origin) 
is dual to the system of coincident $N$ D1-branes at finite temperature. 
The temperature $T$ is inverse of the $\beta\equiv L_y$. 
By taking T-duality along the spatial circle ($x$-direction), 
we obtain the system of $N$ D0-branes, sitting in the compactified 
spatial dimension of radius $(2\pi)^2/L_x$. 
The Wilson line phases correspond to the positions of the D0-branes 
along $x$-direction. 

According to the gauge/gravity duality conjecture~\cite{IMSY98}, 
at low temperature this system is well approximated by type II supergravity, 
and for fixed $\beta$ 
there is a first-order phase transition associated with 
the breakdown of the ${\mathbb Z}_N$ symmetry along $x$-direction~%
\cite{GL_dual gauge theory,AMMPRW05}. 
Below the critical value of the radius $L_{x,c}\sim\sqrt{\beta}$,  
${\mathbb Z}_N$ is broken. Along the temporal direction, ${\mathbb Z}_N$ 
is conjectured to be broken at any nonzero temperature, that is, the system is always 
deconfined. 
For 1d system obtained by dimensional reduction from 2d
$\mathcal{N}=(8,8)$ SYM
it has been confirmed by using the Monte-Carlo simulation~%
\cite{AHNT07,HMNT08,CW08}. 

The physical interpretation of the ${\mathbb Z}_N$ breakdown in string theory is simple. 
In ${\mathbb Z}_N$ unbroken phase, D0-branes (Wilson line phases) 
fill the compact direction uniformly. 
It is a ``uniform black string''\footnote{Do not confuse with ``black
1-brane'' which is a classical solution to type IIB supergravity
corresponding 
to a bunch of D1-branes, while the black string is a solution to type
IIA. 
The black string and the black 1-brane are related by
the T-duality. }.  
If ${\mathbb Z}_N$ is broken completely, 
D0-branes clump to a small region along compact direction;  
it is similar to the usual ``black hole'' (black 0-brane). 
This black hole is small at low temperature, 
and cannot wind on the spatial direction. At high temperature the black hole becomes large, 
and gradually fills the compact dimension, wraps on it and becomes the black string.    
At intermediate temperature the ``nonuniform black string'' may exist. 
In this phase, ${\mathbb Z}_N$ is broken and phases of the Wilson line is distributed 
nonuniformly, but still the density is nonzero everywhere.  

In the high-temperature limit, fermions decouple and the system reduces
to 1d bosonic Yang-Mills. In~\cite{Kawahara:2007fn}
this limit was studied by the
Monte-Carlo simulation and  it was found that the nonuniform black
string phase exists indeed. By fixing the temperature and varying the
size of the spatial circle, there are two phase transitions, namely
between uniform and nonuniform black strings and between the nonuniform
black string and the black hole. The orders of the transitions are of
third and second order, respectively. 

The situation is similar for $(p+1)$-dimensional SYM with 16 supercharges, 
in which the bound state is dual to a bunch of D$p$-branes. 
At low temperature we can analyze it using type II supergravity and 
$({\mathbb Z}_N)^p$ symmetry broken/unbroken phases correspond to 
smeared D0-brane solution and black 0-brane solution. 
For detailed phase structure, see~\cite{HN07}. 

We emphasize again 
that the bound state of scalar eigenvalues is necessary for the black brane description.
In order to understand this statement, let us remind that the black brane is 
the bound state of very large number of D-branes (indeed the gravity picture 
is valid in the large-$N$ limit, where $N$ is nothing but the number of D-branes). 
Because the scalar fields represent the collective 
coordinates of the D-branes, the bound state of the D-branes is exactly
the bound state of scalar eigenvalues.

\subsection{$\mathbb{Z}_N$ breakdown and volume (in)dependence}
Before concluding this section, 
we argue general features of the large-$N$ volume independence. 

In the large-$N$ limit, if the ${\mathbb Z}_N$ symmetry is not broken 
then the system is volume independent (Eguchi-Kawai equivalence)~%
\cite{EK82,NN03}. That is, physical quantities 
do not depend on the volume up to a trivial proportionality factor. 
For example, the free energy per unit volume is volume-independent. 
This property is practically very useful when one studies the large-$N$ 
field theories numerically, because by using small-volume (sometimes zero-volume) 
lattice one can save computational cost. 
Recently its application for a study of the ${\mathbb Z}_N$ unbroken phase of 
4d ${\cal N}=1$ pure SYM~\cite{KUY07}   
has been discussed~\cite{Bringoltz09,BBCS09}. 
In the case of 2d $\mathcal{N}=(2,2)$, however, the ${\mathbb Z}_N$ symmetry is broken 
(although we cannot prove the non-existence of the ${\mathbb Z}_N$ unbroken phase) and 
hence the theory is volume dependent. 

There is another formulation utilizing the Eguchi-Kawai equivalence~\cite{IIST08}. 
In this construction, matrix quantum mechanics around a certain background is 
equivalent to 4d gauge theory on $\mathbb{R}\times S^3$. This theory is manifestly 
volume-dependent because the curvature of the sphere emerges as a parameter. 
This technique can be used to formulate 4d ${\cal N}=1$ SYM~\cite{HMM09}. 
Similarly, 3d gauge theory on $\mathbb{R}\times S^2$ can be formulated by expanding 
the matrix quantum mechanics around a fuzzy sphere and then taking the commutative limit. 
We can also formulate 3d theory on $S^3$ and 2d theory on $S^2$ by using 
a zero-dimensional matrix model.   
These models are presumably corresponding to the ${\mathbb Z}_N$ broken phase. 
\section{Numerical results and the phase structure}\label{sec:simulation}
In this section we study the phase structure of 2d $\mathcal{N}=(2,2)$
SYM in the case of P-P boundary condition numerically. 
Because of the limitation of the resources, 
we study only the case that $L_x=L_y=L$. 

\subsection{The existence of the bound state}\label{sec:bound state}
\begin{figure}[tbp]
\begin{center}
\includegraphics*{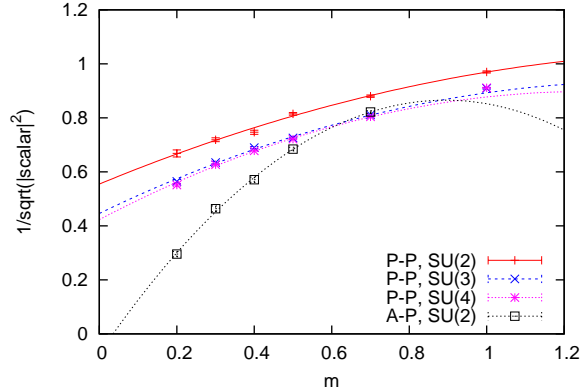}
\caption{ The mass-dependence of scalar eigenvalues at $L=0.707$ is plotted. 
The vertical axis represents 
$\left(\left\langle\sqrt{\frac{1}{NL^2}\int dx\int dy\,\Tr X_i^2}\right\rangle\right)^{-1}$, 
which is the inverse of the typical value of the scalar eigenvalues. 
In P-P boundary condition the scalar eigenvalues remain finite in the $m\to 0$ limit, 
while it diverges in A-P boundary condition. 
Fitting lines are quadratic polynomials of $m$. 
}\label{fig:scalar_norm}
\end{center}
\end{figure}

As we have explained, the system should have a bound state 
in which scalar eigenvalues clump around the origin. 
Although this is stable at large-$N$, for small $N$ it is at most metastable; 
it often collapses and eigenvalues spread along the flat direction.  
In order for the stable simulation, we add the mass term for scalars 
to regularize the flat direction,    
\begin{equation}
S
=
N\int_0^{L} dx\int_0^{L} dy\ 
\Tr\left(
\frac{1}{4}F_{\mu\nu}^2
+
\frac{1}{2}(D_\mu X^i)^2
-
\frac{1}{4}[X^i,X^j]^2
+
m^2 (X^i)^2
\right)
+
(\mbox{fermionic\ part}), 
\end{equation}
and gradually remove it. 
Furthermore we impose the periodic-periodic boundary condition, 
because the metastable state becomes more stable since
the fermion zero-modes provide an attractive force between 
eigenvalues~\cite{AIKKT98}.  
In Fig.~\ref{fig:scalar_norm}, the mass-dependence of 
$\left(\left\langle\sqrt{\frac{1}{NL^2}\int dx\int dy\,\Tr X_i^2}\right\rangle\right)^{-1}$ 
is plotted. (The physical volume is $L=0.707$.) 
As we can see,  it  
converges to a nonzero value, which suggests the existence of the bound state. 
As we will show in \S~\ref{sec:Wilson loop}, 
the mass-dependence of the Wilson loops disappears at $m\lesssim 0.4$. 
This suggests that the mass term is so small   
that it is effective only when the eigenvalue of the scalar fields 
deviates significantly from the typical size of the bound state. 
Furthermore, as we will show in \S~\ref{sec:0d limit}, this state 
corresponds to the bound state in the zero-dimensional matrix model. 
Therefore we conclude that we have constructed the bound state 
mentioned in the previous section. 
We also plot the same quantity with A-P boundary condition. 
It is clear that the scalar eigenvalues diverges in the $m\to 0$ limit. 
Hence this phase is not the bound state. 
Note that the unbounded state exists also with P-P boundary condition.
Indeed, we observed that the norm of the scalar fields sometimes blows up
if we set $m=0$.  This can be interpreted that a meta-stable bound state 
decayed into an unbounded state.
For A-P boundary condition, 
we expect that larger value of $N$ is needed to stabilize the bound
state, as in 1d theory studied in~\cite{AHNT07}.

A few remarks on the previous numerical simulations for $SU(2)$ gauge group~%
\cite{KS08sono2,Suzuki07,KSS07,KS08,Kanamori09} are in order here. 
In~\cite{KS08sono2,KS08,Kanamori09} the same regularization by introduction of 
the mass term is used. 
In these works the scalars seems to diverge in $m\to 0$ limit 
(it is the same as the one with A-P boundary condition, shown in 
Fig.~\ref{fig:scalar_norm}.
See also 
\cite{Kanamori:2008vi}.) 
and hence it is plausible that 
the extrapolation to $m=0$ picks up the different phase from 
the bound state\footnote{
However, the configurations may have a large overlap with the bound
state as well since the scalars are gathered around the origin by the
effect of the finite mass. }. 
It is plausible that this phase corresponds to the ${\mathbb Z}_N$-unbroken phase
discussed in~\cite{KUY07,HMNR09}. 
On the other hand, the phase studied in~\cite{Suzuki07,KSS07} 
should be the bound 
state, because in these papers 
the configurations have been made by
using
the bosonic part of the action and the effect of the fermionic
part has been taken into account by the reweighting method. 
(Note that in the bosonic model the flat direction is 
lifted by quantum corrections and hence only bound states exist.) 
\subsection{Large-$N$ behavior of the Wilson loop and 
breakdown of ${\mathbb Z}_N$ symmetry}\label{sec:Wilson loop}

In this subsection, we study the Wilson loop $\Tr W_\mu/N$, where 
$W_\mu$ is a unitary matrix which is defined by 
\begin{eqnarray}
W_x(y)
=
P\exp\left(
i\int_0^L dx A_x(x,y)
\right), 
\qquad
W_y(x)
=
P\exp\left(
i\int_0^L dy A_y(x,y)
\right). 
\end{eqnarray}
On lattice, it can be obtained by multiplying the link variables. 
Because of the translational invariance of the model, the expectation value 
does not depend on the coordinate. 
 
At finite $N$, there is a tunneling between different vacua related by a 
multiplication of a phase factor (\ref{eq:phase-factor}) and hence the
expectation value
$\langle\frac{1}{N} \Tr W_\mu\rangle$ is zero. (In the large-$N$ limit, the
tunneling is suppressed and the expectation value can be non-zero.)
Therefore, in order to see a possible 
symmetry breakdown at large-$N$,  
we fix the $({\mathbb Z}_N)^2$ symmetry so that 
${\rm Re}\sum_{j=1}^k
\Tr W_\mu^{(j)}$ becomes maximum.  
That is, $W_\mu^{(j)}$ is replaced by $e^{2\pi i n_\mu/N} W_\mu^{(j)}$
with a suitable $n_\mu$ which satisfies the above condition.
In our numerical simulation, we have performed this ${\mathbb Z}_N$ fixing 
at each measurement.  

\begin{figure}[tbp]
\begin{center}
\includegraphics*{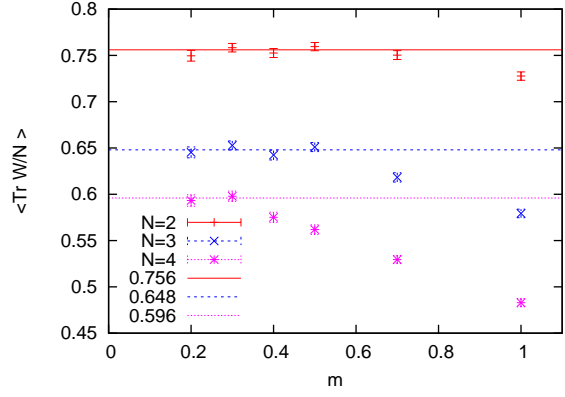}
\caption{ The expectation value of the Wilson loop $\langle \Tr W/N \rangle$
at $L=0.707$ is plotted 
for $0\le m\le 1$. At small values of $m$, the mass-dependence disappears.  
}\label{fig:mass_dependence}
\end{center}
\end{figure}

Below we show the expectation values of the Wilson loops. 
We found that $\langle\frac{1}{N} \Tr W_x\rangle$ and $\langle\frac{1}{N} \Tr W_y\rangle$ are 
the same in the error. Therefore we use the average of them
$\langle\frac{1}{N} \Tr W \rangle$, where $W\equiv (W_x+W_y)/2$.
(In the bosonic models, they take different values for some 
parameters~\cite{NN03,HN07}.) 

In Fig.~\ref{fig:mass_dependence}, the mass-dependence of 
the Wilson loop is plotted. (The size of the torus is $L=0.707$.)
The mass-dependence disappears at small values of $m$, 
which allows us to use small fixed value of $m$.
In practice, we use $m=0.2$, $0.3$ and $0.4$.%
\footnote{
Without introducing the mass $m$, scalar eigenvalues stays around the origin 
for a while and then go to infinity. It is consistent with the interpretation 
that the bound state is metastable. It is not impossible to evaluate expectation values  
by using only configurations from metastable configuration, although the error 
is rather large and the result is less reliable. 
The result is consistent with the constant fitting shown in  
Fig.~\ref{fig:mass_dependence}. 
}

\begin{figure}[tbp]
\begin{center}
\includegraphics*{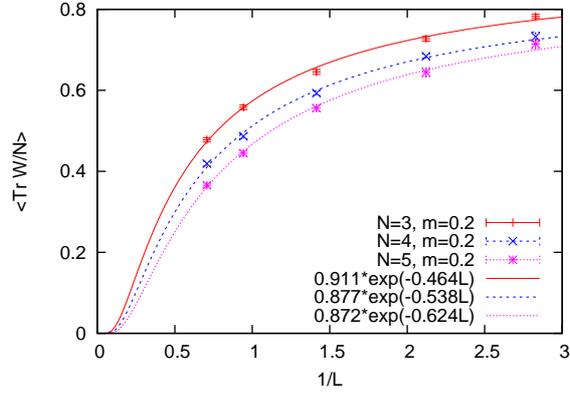}
\caption{ The expectation value of the Wilson loop $\langle \Tr W/N \rangle$
for $N=3,4$ and $5$. 
Mass parameter is $m=0.2$.  
}\label{fig:WilsonLoops}
\end{center}
\end{figure}
\begin{figure}[tbp]
\begin{center}
\includegraphics*{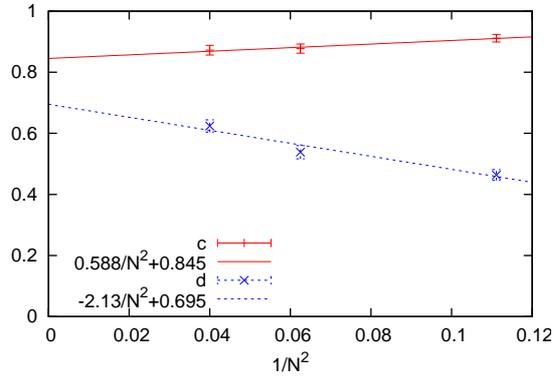}
\caption{$N$-dependence of the coefficients in (\ref{fitting_wilson}), 
$N=3,4$ and $5$. 
Mass parameter is $m=0.2$.  
}\label{fig:WilsonCoeff}
\end{center}
\end{figure}
In Fig.~\ref{fig:WilsonLoops}, we plot the expectation value of the Wilson loop against 
the size of the torus $L$. For each $N$, the expectation value can be fitted by 
\begin{eqnarray}
\left\langle\frac{1}{N}\Tr W \right\rangle 
=
c\times\exp\left(
-d\cdot L
\right), 
\label{fitting_wilson}
\end{eqnarray}
where $c$ and $d$ are real and positive.  
By using the data at 
$L=0.471$, $0.707$, $1.061$, $1.414$ and $m=0.2$, 
we obtain $c=0.911\pm 0.012, d=0.464\pm 0.018$ for $N=3$, 
$c=0.877\pm 0.015, d=0.538\pm 0.022$ for $N=4$ and 
$c=0.872\pm 0.016, d=0.624\pm 0.020$ for $N=5$. 
Extrapolating these coefficients to $N=\infty$, we obtain                                        
$c=0.845\pm 0.009$ and $d=0.695\pm 0.040$ (Fig.~\ref{fig:WilsonCoeff}). 
Therefore, in the large-$N$ limit, 
the expectation value of the Wilson loop is nonzero at finite volume, 
and hence the ${\mathbb Z}_N$ symmetry is broken.  
Note that the ${\mathbb Z}_N$ symmetry is restored in the large volume limit as expected. 
\subsection{Zero-volume limit}\label{sec:0d limit}

As we have seen above, the simulation data suggests that 
the ${\mathbb Z}_N$ symmetry is broken completely in the zero-volume limit, 
that is, the expectation value of the Wilson loop  
$\left\langle\frac{1}{N}\Tr W_\mu\right\rangle$ 
becomes close to $1$.
Then the model reduces to the one-point reduction of 4d ${\cal N}=1$ SYM\footnote{
Unless the ${\mathbb Z}_N$ symmetry breaks completely, the zero-volume limit 
is different from the naive dimensionally reduced model; see~\cite{AMMPRW05}. 
}, which has been studied numerically in~\cite{AABHN00}. 
Hence, by comparing the 2d and 0d models, 
we can check the validity of our simulation. 

Let us start with the action in two dimensions, 
\begin{eqnarray}
S_{2d}
=
N
\int_0^{L} dx
\int_0^{L} dy
\, \Tr\left(
\frac{1}{2}(D_\mu Y_i)^2
-\frac{1}{4}[Y_i,Y_j]^2
+
\frac{1}{4}F_{\mu\nu}^2
\right). 
\end{eqnarray} 
Here we denoted the scalar fields as $Y_i$ in order to distinguish them 
from the scalars $X_I$ in the zero-dimensional model. 
(For simplicity, we write down only the bosonic part.) 
By neglecting nonzero-modes, this action reduces to the zero-dimensional one. 
By using  
$X_I=(X_\mu, X_i)$ defined by 
\begin{eqnarray}
X_i
=
L^{-3/2}\int_0^L dx\int_0^L dy\, Y_i, 
\quad
X_\mu=
L^{-3/2}\int_0^L dx\int_0^L dy\, A_\mu, 
\end{eqnarray} 
we obtain the zero-dimensional model with the canonical normalization, 
\begin{eqnarray}
S_{0d}
=
-\frac{N}{4}
\Tr
[X_I,X_J]^2. 
\end{eqnarray} 
Here, the gauge field $A_\mu$ can be obtained by 
\begin{eqnarray}
\frac{1}{2}\left[
-i(W_\mu-1)+c.c.
\right]
\simeq
\int_0^L dx^\mu A_\mu, 
\label{eq:Wilson loop and gauge field}
\end{eqnarray}
where in the r.h.s.\ the index $\mu$ is not contracted.  
When we consider the $k\times k$ square lattice, 
we can calculate $k$ Wilson loops along both $x$ and $y$ directions, 
$W_\mu^{(1)},\cdots,W_\mu^{(k)}$. $X_\mu$ can be obtained as\footnote{
Because $X_\mu$ should be traceless, we have to 
fix the $({\mathbb Z}_N)^2$ symmetry so that 
${\rm Re} \sum_{j=1}^k
\Tr W_\mu^{(j)}$ becomes maximum, or equivalently the argument of 
$\sum_{j=1}^k \Tr W_\mu^{(j)}$ is minimum. 
(It is the same as the ${\mathbb Z}_N$ fixing performed in \S~\ref{sec:Wilson loop}.)
Equations (\ref{eq:Wilson loop and gauge field}) and 
(\ref{eq:Wilson loop and scalar field in 0d}) hold 
only with this condition. 
In our numerical simulation, we have performed this ${\mathbb Z}_N$ fixing 
at each measurement.  
}  

\begin{eqnarray}
X_\mu
\simeq
\frac{1}{k\sqrt{L}}\sum_{j=1}^k
\frac{1}{2}\left[
-i(W_\mu^{(j)}-1)+h.c.
\right]. 
\label{eq:Wilson loop and scalar field in 0d}
\end{eqnarray}  

There is a subtlety in the discussion above -- 
``zero-mode'' is not a gauge-invariant notion. 
(A constant field configuration, which corresponds to the zero-dimensional model, 
can be transformed to rapidly varying configuration just by a gauge transformation!)   
Therefore, we have to choose a suitable gauge 
in which zero modes reproduce the zero-dimensional matrix model~\cite{AHHS09}. 
In the current setup, we should maximize (resp. minimize) 
the zero-mode (resp. nonzero-mode) contributions, so that the configuration 
is as static as possible.  
For that purpose we choose the gauge so that $\Tr(X_i^2+X_\mu^2)$ is maximum. 

\begin{figure}[htbp]
 \hfil
 \begin{minipage}{0.45\hsize}
   \begin{center}
   \includegraphics[width=0.96\linewidth]{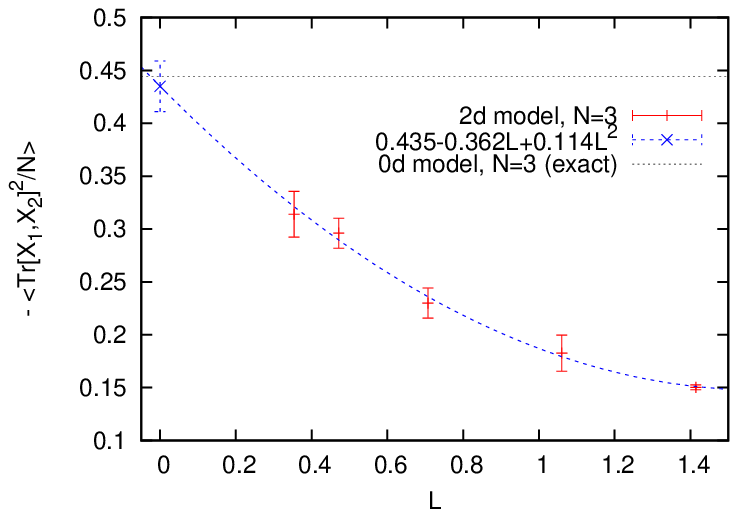}
  \end{center}
\caption{ 
$\left\langle-\frac{1}{N}\Tr[X_1,X_2]^2\right\rangle$, 
$N=3$, calculated from gauge fields.  
}\label{fig:TrF2_gauge_SU3}
 \end{minipage}
 \hfil
 \begin{minipage}{0.45\hsize}
   \begin{center}
   \includegraphics[width=.95\linewidth]{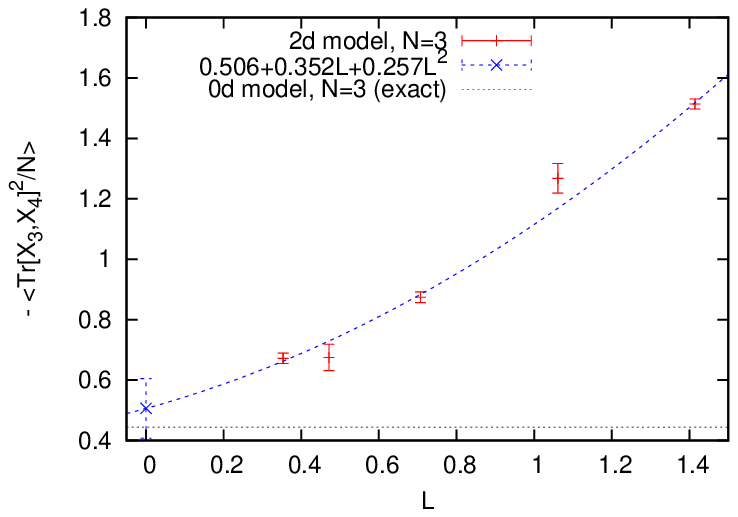}
  \end{center}
\caption{ 
$\left\langle-\frac{1}{N}\Tr[X_3,X_4]^2\right\rangle$, 
$N=3$, calculated from scalar fields.  
}\label{fig:TrF2_scalar_SU3}
 \end{minipage}
\end{figure}

\begin{figure}[htbp]
 \hfil
 \begin{minipage}{0.45\hsize}
   \begin{center}
   \includegraphics[
    width=0.96\linewidth
    ]{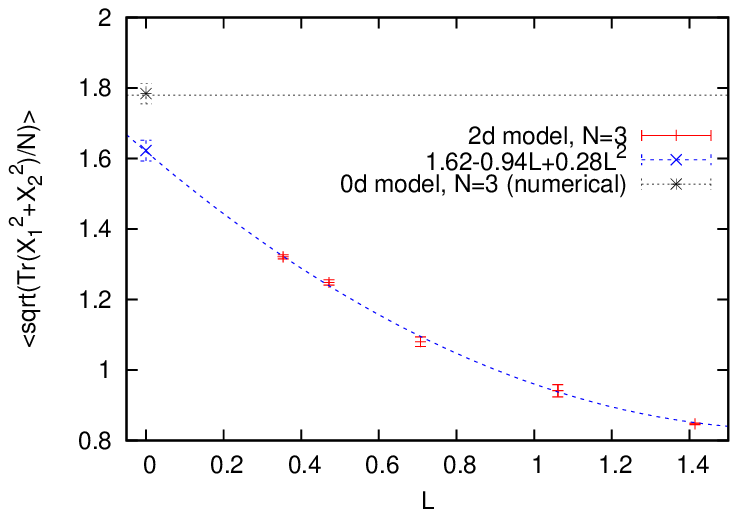}
  \end{center}
\caption{ 
$
\left\langle 
\sqrt{ 
\frac{1}{N}\Tr
\left(X_1^2+X_2^2\right)}
\right\rangle$, 
$N=3$, calculated from gauge fields.   
}\label{fig:TrX2_gauge_SU3}
 \end{minipage}
\hfil
 \begin{minipage}{0.45\hsize}
   \begin{center}
   \includegraphics[
    width=0.96\linewidth
    ]{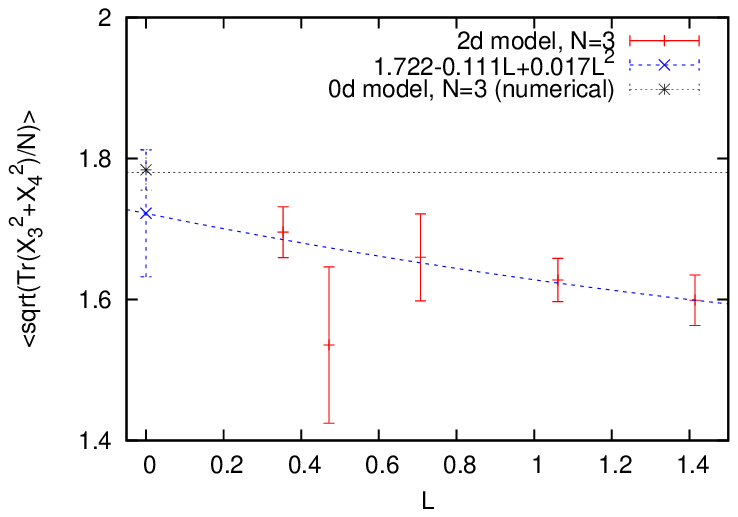}
  \end{center}
\caption{ 
$
\left\langle 
\sqrt{ 
\frac{1}{N}\Tr
\left(X_3^2+X_4^2\right)}
\right\rangle$, $N=3$, 
 calculated from scalar fields.  
}\label{fig:TrX2_scalar_SU3}
 \end{minipage}
\end{figure}

In this gauge we can compare the small volume behavior with 
zero-dimensional matrix model. 
The simplest quantity is 
$\left\langle -\frac{1}{N}\Tr[X_I,X_J]^2\right\rangle$, 
which can be evaluated exactly~\cite{HNT98}:
\begin{eqnarray}
\left\langle 
-\frac{1}{N}
\Tr[X_1,X_2]^2
\right\rangle
=
\left\langle 
-\frac{1}{N}
\Tr[X_3,X_4]^2
\right\rangle
=
\frac{1}{2}
\left(
1-\frac{1}{N^2}
\right). 
\end{eqnarray}
In Fig.~\ref{fig:TrF2_gauge_SU3} and Fig.~\ref{fig:TrF2_scalar_SU3},  
the corresponding quantity in two-dimensional SYM is plotted.  
In Fig.~\ref{fig:TrX2_gauge_SU3} and Fig.~\ref{fig:TrX2_scalar_SU3},  
$\left\langle\sqrt{\sum_{\mu=1}^2\frac{1}{N}\Tr X_\mu^2}\right\rangle$ 
and the corresponding quantity in two dimensions are plotted. 
In these plots, the results at $m=0$ is obtained by extrapolating 
results at $m=0.2,0.3$ and $m=0.4$ by a straight line. 
Then we have fitted them by a polynomial $A+BL+CL^2$.  
We have evaluated 
$\left\langle\sqrt{\sum_{\mu=1}^2\frac{1}{N}\Tr X_\mu^2}\right\rangle$ 
numerically in 0d theory and 
found that it is $1.78\pm 0.03$ for $SU(3)$.  
In Fig.~\ref{fig:TrF2_gauge_SU3}, Fig.~\ref{fig:TrF2_scalar_SU3} 
and Fig.~\ref{fig:TrX2_scalar_SU3}
the data is consistent with 0d model results for $SU(3)$.
(In Fig.~\ref{fig:TrX2_gauge_SU3}, the extrapolated value differs slightly   
from the 0d result. One possible reason is this quantity is sensitive to the error 
in the approximation in (\ref{eq:Wilson loop and scalar field in 0d}), which is 
exact in the limit $W_\mu=\textbf{1}_N$.) 
Since our 2d simulation is smoothly connected to the 0d model,
we conclude that the bound state we have constructed is 
exactly the one discussed in~\S \ref{subsec:2dSYM}.
\subsection{Supersymmetry}

\begin{figure}[tbp]
\begin{center}
\includegraphics*{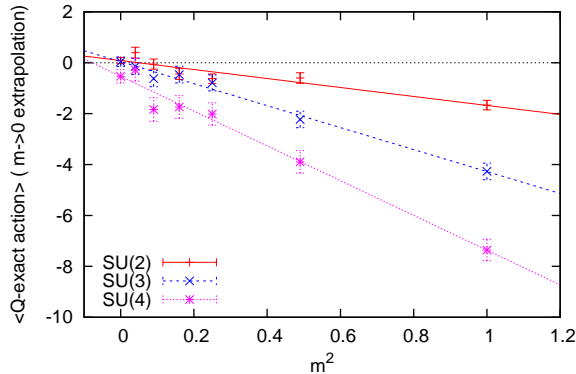}
\caption{Mass-dependence of the action at $L=0.707$. 
The points at $m=0$ are obtained by extrapolations.  
In the $m\to 0$ limit, where the continuum theory becomes supersymmetric, 
the expectation value is consistent with zero ( SU(4) case seems
 nonzero but consistent with zero in 2 standard deviations).      
}\label{fig:action}
\end{center}
\end{figure}

It is important to study whether the bound state preserves supersymmetry or not. 
Here we show the expectation value of the action. 
In the Sugino model the action is of the form 
\begin{eqnarray}
S=\{Q,{\cal O}\}, 
\end{eqnarray}
where $Q$ is one of four supercharges which is exactly kept in the regularization. 
Therefore, the expectation value of the action must be zero if the vacuum 
is invariant under the supersymmetry generated by $Q$. \footnote{
Because we are picking up fluctuations around one specific state, 
$Q$-exact quantities can have nonzero expectation values 
without introducing any external fields nor temperature.  
Remember that we are studying the bound state only while there is
an unbounded state as well.
} As we can see from Fig.~\ref{fig:action}, the expectation value of the action is 
consistent with zero. 

Strictly speaking, there is a subtlety for studying the breakdown of the supersymmetry 
with periodic boundary conditions. The reason is as follows~\cite{KSS07}. 
In the simulation, 
we obtain the expectation values normalized as 
\begin{eqnarray}
\frac{\langle 0|\{Q,{\cal O}\}|0\rangle}{\langle 0|0\rangle}
\end{eqnarray}
where the denominator $\langle 0|0\rangle$ is the partition function. 
In order for the simulation to make sense, the denominator  
must be nonzero, but this condition can be broken. Indeed, if the continuum 
spectrum is absent and the Witten index is well-defined, then the partition function 
with the periodic boundary conditions is nothing but the Witten index, which is zero 
if the supersymmetry is spontaneously broken. 
In the present case, because the continuum spectrum exists due to the
existence of the unbounded state 
and there is an ambiguity for the definition of the Witten index, 
the above argument cannot be applied straightforwardly.   
Although it is difficult to exclude the possibility that the partition function is zero, 
we believe the partition function is nonzero 
because this system does not suffer from the sign problem. 
Clarification of this point is desirable. 

In order to see whether the supersymmetry is spontaneously 
broken or not, the simplest and unambiguous 
way is to put the theory at finite temperature and calculate the energy density~\cite{KSS07}. 
Supersymmetry is not broken if and only if the energy density is zero at zero temperature. 
It is an important future problem to study 2d theory at large-$N$ and at finite temperature,  
and confirm that the energy converges to zero\footnote{ 
For matrix quantum mechanics which is obtained from 2d ${\cal N}=(2,2)$ SYM 
through the dimensional reduction, Smilga conjectured that the supersymmetry is broken 
in the bound state phase~\cite{Smilga08}. 
In the unbounded state, the supersymmetry is argued to be unbroken. 
The corresponding phase in 2d theory has been studied in~\cite{KS08,KS08sono2,Kanamori09} 
and the ground state energy was found to be consistent with zero~\cite{Kanamori09}. 
}.

As we will see in \S~\ref{sec:continuum limit}, 
it is plausible that the explicit supersymmetry breaking lattice artifacts 
disappears in the continuum limit. 
Then, the fact that the expectation value of the action is zero suggests the absence
of the spontaneous supersymmetry breaking. 
It is desirable to check it more rigorously by measuring the energy 
of the system.

\subsection{Convergence to the continuum limit}\label{sec:continuum limit}

In this subsection, we show that the lattice spacing used in the current 
simulation is small enough to study the system quantitatively.

\begin{table}
 \hfil
 \begin{tabular}{c|c|c}
  gauge group & volume & lattice size\\
  \hline
  $SU(2)$ & 0.707 & $4\times 4$\\     
  $SU(2)$ & 1.414 & $8\times 8,\ 6\times 6,\ 5\times 5,\ 4\times 4$\\     
  \hline
  $SU(3)$ & 0.354, 0.471, 0.707 & $4\times 4$\\    
  $SU(3)$ & 1.061, 1.414 & $6\times 6$\\  
  \hline
  $SU(4)$ & 0.354, 0.471, 0.707 & $4\times 4$\\    
  $SU(4)$ & 1.061, 1.414 & $6\times 6$\\  
  \hline
  $SU(5)$ & 0.354, 0.471, 0.707, 1.061, 1.414 & $4\times 4$\\    
 \end{tabular}
 \caption{ A list of lattice sizes used in this work. 
}\label{table:lattice size}
\end{table}

\begin{figure}[tbp]
\begin{center}
\includegraphics{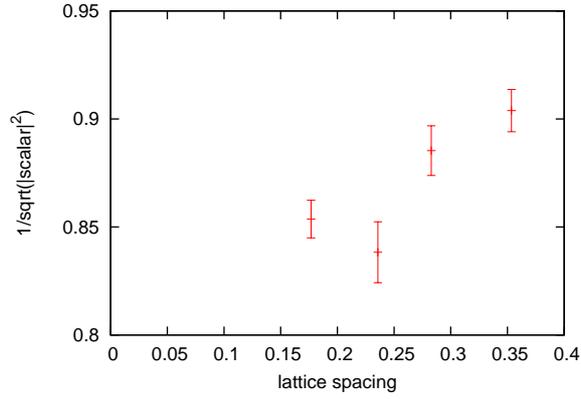}
\caption{Dependence of the inverse of the scalar norm 
$\left(\left\langle\sqrt{\frac{1}{NL^2}\int dx\int dy\,\Tr X_i^2}\right\rangle\right)^{-1}$ 
on the lattice spacing at $N=2, L=1.414, m=0.2$.   
}\label{fig:scalar_latticespacing_dep}
\end{center}
\end{figure}

\begin{figure}[tbp]
\begin{center}
\includegraphics*{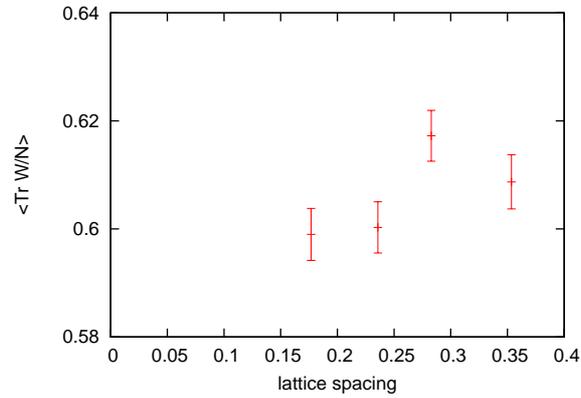}
\caption{Dependence of the Wilson loop to the lattice spacing at $N=2, L=1.414, m=0.2$.   
}\label{fig:Wilson_latticespacing_dep}
\end{center}
\end{figure}

\begin{figure}[tbp]
\begin{center}
\includegraphics{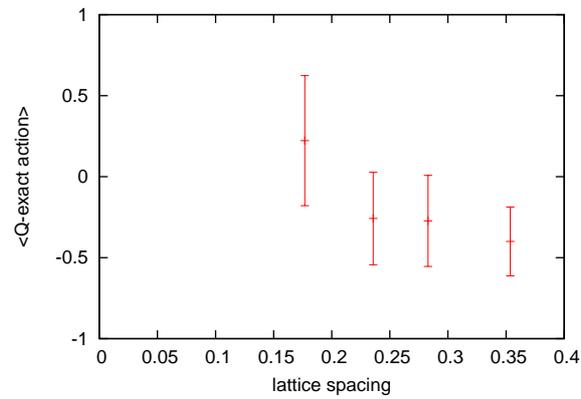}
\caption{Dependence of the action on the lattice spacing at $N=2, L=1.414, m=0.2$.   
}\label{fig:action_latticespacing_dep}
\end{center}
\end{figure}

The lattice sizes used in this work are listed in Table.~\ref{table:lattice size}. 
In Fig.~\ref{fig:scalar_latticespacing_dep}, 
Fig.~\ref{fig:Wilson_latticespacing_dep} and 
Fig.~\ref{fig:action_latticespacing_dep}, 
we plot the lattice spacing dependences of 
the extent of the scalar eigenvalues, the Wilson loop 
and the action. 
Lattice size was taken to be  
$4\times 4$,  $5\times 5$, $6\times 6$ and $8\times 8$ and other parameters are taken to be 
$N=2, L=1.414, m=0.2$.
It turns out that the expectation values, especially that of the Wilson loop,  
are not sensitive to the lattice spacing used in the simulation. 
Note that $4\times 4$ lattice at $L=1.414$ is the coarsest one in this work. 
Therefore the numerical data used in the previous sections is sufficiently close to the continuum limit.

\begin{figure}
 \hfil
 \includegraphics[
 ]{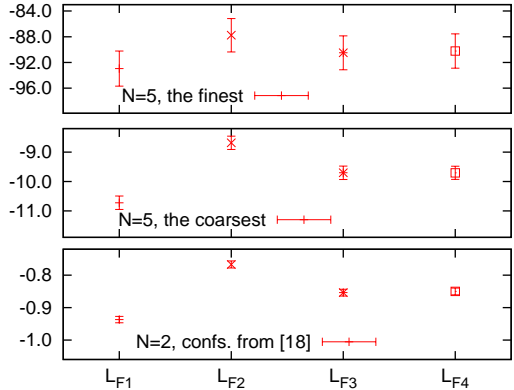}
 \caption{Expectation values of $\mathcal{L}_{Fi} (i=1,...,4)$, 
the almost degenerate behavior
 implies the that the simulation is close to the continuum limit.
The lowest plot uses the same configuration used to plot Fig.~8 in 
\cite{KS08sono2}.}
 \label{fig:lf}
\end{figure}

In order to see that the supersymmetries which are broken by a lattice artifact 
are restored in the continuum limit, 
we utilize the discrete symmetry of the system which is related to the R-symmetry. 
The action in the continuum has the following discrete symmetries
(we follow the notation in the Appendix):
\begin{align}
 \frac{\eta}{2}&\to \chi, & \chi&\to -\frac{\eta}{2}, &
   \psi_0 &\to \psi_1, & \psi_1 &\to -\psi_0
 \label{eq:discretesym1}
\end{align}
and
\begin{align}
 \frac{\eta}{2}&\to \psi_1, & \chi&\to\psi_0,
 &\psi_1&\to\frac{\eta}{2}, & \psi_0&\to\chi, 
 &\phi &\to -\bar{\phi}, & \bar{\phi}&\to -\phi.
 \label{eq:discretesym2}
\end{align} 
Because of these symmetries, in the continuum limit, all four
Yukawa interaction terms $\mathcal{L}_{F1},...,\mathcal{L}_{F4}$
(for explicit forms, see Appendix)
should give an identical expectation value.
In Fig.~\ref{fig:lf}, we plot them for the coarsest ($L=1.414$
with $4\times 4$ lattice) and the finest  
($L=0.354$ with $4\times 4$ lattice) cases with $N=5$. 
\footnote{
Because we use a unit  `t Hooft coupling $\lambda=N g^2=1$,
and because of a difference of the normalization of the kinetic term, 
the lattice spacing defined by the same manner as~\cite{KS08sono2} 
becomes smaller by a factor $\sqrt{N/2}$.
}
We also plot $N=2$ case using the same configuration used to plot
Fig.~8 in~\cite{KS08sono2}. 
The plots show that the expectation values from the finest lattice  
are almost degenerated
for all $\mathcal{L}_{Fi}$ ($i=1,...4$) and it strongly suggests 
that this discrete symmetry is restored in the continuum limit.  
Therefore we can expect that supersymmetries broken by a lattice artifact 
are restored in the continuum limit, because they are related to 
the exactly kept supersymmetry via this discrete symmetry. 
Note that, although the coarsest case would have some effects 
from the lattice artifacts, at least for the quantities like the Wilson loop 
such effects are negligible, as we have seen above. 

A remark on the previous simulation~\cite{KS08sono2} is in order here.  
In~\cite{KS08sono2}, the restoration of supersymmetries which are 
explicitly broken by lattice artifacts  
has been studied. In that work, expectation values 
of two-point functions are used.   
(In the present simulation, this technique cannot be used because the lattice is too small 
to calculate the two-point functions.)  
In Fig.~8 and~9 in~\cite{KS08sono2}, four different two-point functions are plotted
for P-P case which should be degenerated because of the above discrete symmetries.
The plots, however, does not show the degenerate behavior and hence 
it seems that the simulation is far from the continuum limit.\footnote{
Because of this and noisy results for 
partially conserved supercurrent relation, \cite{KS08sono2} did not study
P-P case extensively.
} 
The non-degenerate behavior is again found in the lowest plot in
Fig.~\ref{fig:lf} while in the first plot from the current simulation 
almost degenerated behavior can be seen. 

\section{Conclusion and discussions}\label{sec:conclusion}
In this paper we have studied large-$N$ properties of two-dimensional 
${\cal N}=(2,2)$ SYM. Especially we have established the existence of 
the bound state in which scalar eigenvalues clump around the origin. 
It makes the simulation well-defined in spite of the existence of 
the flat direction along which scalar eigenvalues spread to infinity.  
We also have shown numerically that at finite volume global $(\mathbb Z_N)^2$ is broken.
This symmetry is restored in the large volume limit.

If the $({\mathbb Z}_N)^2$ symmetry were not broken, 
then because of the Eguchi-Kawai equivalence 
the expectation values of the loops would not depend on 
the volume in the large-$N$ limit~\cite{EK82,NN03}.  
The Eguchi-Kawai equivalence works for bosonic Yang-Mills 
only above a critical volume~\cite{NN03} 
because the $({\mathbb Z}_N)^4$ symmetry is broken below it. 
The twisted~\cite{TEK} and quenched~\cite{QEK} Eguchi-Kawai models 
were believed to cure this problem, but recently turned out 
to fail in the large-$N$ limit~\cite{TEKbreakdown,QEKbreakdown}. 
The reason is that the backgrounds collapses due to large fluctuations. 
(Another deformation has been proposed in~\cite{DEK}.) 
For supersymmetric theories the ${\mathbb Z}_N$ symmetry was expected not to 
be broken, but as shown in this paper, it is not necessarily the case. 
However, once we combine the idea of twisted or quenched Eguchi-Kawai 
model with the supersymmetry    
then the ${\mathbb Z}_N$-unbroken background becomes 
stable~\cite{AHH08} 
and the Eguchi-Kawai equivalence can hold. 
A concrete proposal has been given in~\cite{IIST08} \footnote{ 
The validity of~\cite{IIST08} is discussed in~\cite{IKNT08}.}.  
Studying this direction further is very important, because 
using the Eguchi-Kawai equivalence we can regularize planar SYM 
in higher dimensions (3d and 4d) 
without using lattice, and hence we may avoid the difficulties in lattice SUSY. 

Although finite-temperature properties of the bound state is very interesting 
in connection to the black hole thermodynamics, 
it is much more difficult to study because we need to take $N$ to be rather large. 
The same difficulty exists in matrix quantum mechanics, 
though large enough $N$ (say $N=16$) can be taken in that case~\cite{AHNT07} 
because the model requires less computational resources. 
In~\cite{CW08}, to avoid this difficulty at rather small $N$, 
finite-temperature properties 
of the matrix quantum mechanics have been studied by performing the simulation 
with the periodic boundary condition and then taking into account 
the effect of the antiperiodic boundary condition by the reweighting method. 
This method would work for the ${\mathbb Z}_N$-broken phase (``black hole'' and 
``non-uniform black string''). However, for the ${\mathbb Z}_N$-unbroken phase, 
this method might not work because we expect a severer overlapping problem.   
In any case, we expect an unambiguous study of the thermal properties 
with large enough $N$ will be possible in near future. 
It will provide us with valuable insights into black hole/black string physics.

\begin{center}
\section*{Acknowledgment}
\end{center}
A part of the program used in this work was developed from
the ones used in collaborations of I.~K. with H.~Suzuki.
The authors would like to thank 
O.~Aharony, B.~Bringoltz, D.~Kadoh, S.~Matsuura, J.~Nishimura, T.~Nishioka, 
D.~Robles-Llana, H.~Shimada, H.~Suzuki, 
M.~\"{U}nsal and L.~Yaffe for discussions. 
The computations were carried out on PC clusters at 
Yukawa Institute and RIKEN RSCC. 
The work of I.~K. was supported by the Nishina Memorial Foundation.

\appendix

\section{Simulation details} 
\subsection{The Sugino model}
We consider the ${\cal N}=(2,2)$ supersymmetric Yang-Mills theory 
on $T^2$, 
whose action is given by 
\begin{align}
\lefteqn{S_{continuum}}\quad \nonumber\\
&=
N\int_0^{L_x} dx\int_0^{L_y}dy
\, \Tr 
\left\{
\frac{1}{4}F_{\mu\nu}^2
+
\frac{1}{2}(D_\mu X_i)^2
-
\frac{1}{4}[X_i,X_j]^2
-
\frac{i}{2}\bar{\psi}\Gamma^\mu D_\mu\psi
-
\frac{1}{2}\bar{\psi}\Gamma^i [X_i,\psi]
\right\}, 
\end{align}
which is obtained from four-dimensional ${\cal N}=1$ SYM through the dimensional reduction.    

As a discretization, we use Sugino's lattice action~\cite{Sugino04}
\footnote{Here we follow the notation in~\cite{Suzuki07, KS08sono2}. 
Under a suitable representation of the Gamma matrices,
the fermion $\psi$ can be taken as
$\psi^T=(\psi_0, \psi_1, \chi, \eta/2)$.
},
\begin{eqnarray}
S_{lattice}
=
a_x a_y\sum_{\vec{x}}
\left\{
\sum_{i=1}^3{\cal L}_{Bi}(\vec{x})
+
\sum_{i=1}^6{\cal L}_{Fi}(\vec{x})
\right\}
+{\rm (auxiliary\ field)}, 
\end{eqnarray}
where 
\begin{align}
{\cal L}_{B1}(\vec{x})
&=
\frac{N}{4a_x^2 a_y^2}
\Tr
[\phi(\vec{x}),\bar{\phi}(\vec{x})]^2, 
\\
{\cal L}_{B2}(\vec{x})
&=
\frac{N}{4a_x^2 a_y^2}
\Tr
\hat{\Phi}_{TL}(\vec{x})^2, 
\\
{\cal L}_{B3}(\vec{x})
&=
\frac{N}{a_x^3 a_y}
\Tr\bigl\{
\left(
\phi(\vec{x})-U_x(\vec{x})\phi(\vec{x}+a_x\hat{x})U_x(\vec{x})^{-1}
\right)
\nonumber\\
& 
\hspace{3cm}\times
\left(
\bar{\phi}(\vec{x})-U_x(\vec{x})\bar{\phi}(\vec{x}+a_x\hat{x})U_x(\vec{x})^{-1}
\right)
\bigl\}
\nonumber\\
& 
+
\frac{N}{a_x a_y^3}
\Tr\bigl\{
\left(
\phi(\vec{x})-U_y(\vec{x})\phi(\vec{x}+a_y\hat{y})U_y(\vec{x})^{-1}
\right)
\nonumber\\
& 
\hspace{3cm}\times
\left(
\bar{\phi}(\vec{x})-U_y(\vec{x})\bar{\phi}(\vec{x}+a_y\hat{y})U_y(\vec{x})^{-1}
\right)
\bigl\}
\end{align}
and
\begin{align}
{\cal L}_{F1}(\vec{x})
&=
-\frac{N}{4a_x^2 a_y^2}
\Tr\left(
\eta(\vec{x})[\phi(\vec{x}),\eta(\vec{x})]
\right), 
\\
{\cal L}_{F2}(\vec{x})
&=
-\frac{N}{a_x^2 a_y^2}
\Tr\left(
\chi(\vec{x})[\phi(\vec{x}),\chi(\vec{x})]
\right), 
\\
{\cal L}_{F3}(\vec{x})
&=
-\frac{N}{a_x^3 a_y}
\Tr\left\{
\psi_0(\vec{x})\psi_0(\vec{x})
\left(
\bar{\phi}(\vec{x})+U_x(\vec{x})\bar{\phi}(\vec{x}+a_x\hat{x})U_x(\vec{x})^{-1}
\right)
\right\}, 
\\
\displaybreak[3]
{\cal L}_{F4}(\vec{x})
&=
-\frac{N}{a_x a_y^3}
\Tr\left\{
\psi_1(\vec{x})\psi_1(\vec{x})
\left(
\bar{\phi}(\vec{x})+U_y(\vec{x})\bar{\phi}(\vec{x}+a_y\hat{y})U_y(\vec{x})^{-1}
\right)
\right\}, 
\\
{\cal L}_{F5}(\vec{x})
&=
i\frac{N}{a_x^2 a_y^2}
\Tr\left(
\chi(\vec{x})\cdot Q\hat{\Phi}(\vec{x})
\right), 
\\
{\cal L}_{F6}(\vec{x})
&=
-i\frac{N}{a_x^3 a_y}
\Tr\bigl\{
\psi_0
\left(
\eta(\vec{x})-U_x(\vec{x})\eta(\vec{x}+a_x\hat{x})U_x(\vec{x})^{-1}
\right)
\bigl\}
\nonumber\\*
& 
-i\frac{N}{a_x a_y^3}
\Tr\bigl\{
\psi_1
\left(
\eta(\vec{x})-U_y(\vec{x})\eta(\vec{x}+a_y\hat{y})U_y(\vec{x})^{-1}
\right)
\bigl\}, 
\end{align}
where $U(\vec{x},\mu)$ are gauge link variables, $\phi(\vec{x})$ is a complex scalar, 
$\eta(\vec{x})$, $\chi(\vec{x})$ and $\psi_\mu(\vec{x})$ are fermion field, 
$a_x$ and $a_y$ are lattice spacings \footnote{In the actual simulation 
we have used the isotropic lattice, $a_x=a_y$.}, 
$\epsilon$ is a real parameter which must be chosen appropriately 
for each $N$ (in this work, we used $\epsilon=2.6$), 
\begin{align}
\hat{\Phi}(\vec{x})
&=
\frac{-i(P(\vec{x})-P(\vec{x})^{-1})}{1-|1-P(\vec{x})|^2/\epsilon^2}, 
\qquad
\hat{\Phi}_{TL}(\vec{x})
=
\hat{\Phi}(\vec{x})
-
\frac{1}{N}\left(\Tr\hat{\Phi}(\vec{x})\right)\cdot\textbf{1}, 
\end{align}
where $P(\vec{x})=U_x(\vec{x})U_y(\vec{x}+\hat{x})U_x^\dagger(\vec{x}+\hat{y})U_y^\dagger(\vec{x})$ is 
the plaquette variable, 
and $Q$ generates one of the four supersymmetries, 
\begin{align}
 QU_\mu(\vec{x}) &= i \psi_\mu(\vec{x})U_\mu(\vec{x}), \\
 Q\psi_\mu(\vec{x})
  &= i\psi_\mu(\vec{x})\psi_\mu(\vec{x})
     -i\bigl( \phi(\vec{x})-U_\mu(\vec{x})\phi(\vec{x}+a_\mu\hat{\mu})U_\mu(\vec{x})^{-1}\bigr), \\
 Q\phi(\vec{x}) &= 0, \\
 Q\chi(\vec{x}) &= H(\vec{x}), \\
 QH(\vec{\vec{x}}) &= [\phi(\vec{x}), \chi(\vec{x})], \\
 Q\phibar(\vec{x}) &= \eta(\vec{x}), \\
 Q\eta(\vec{x}) &= [\phi(\vec{x}), \phibar(\vec{x})].
\end{align}
Sugino's action $S_{lattice}$ is invariant under the supersymmetry 
generated by $Q$, 
because $Q$ is nilpotent up to gauge transformation 
and $S$ can be written in a $Q$-exact form. 

In~\cite{Sugino04}, using super-renormalizability and symmetry argument, 
it was shown that other three supersymmetries, 
which is broken by a lattice artifact at the discretized level, 
is restored in the continuum limit.   
Furthermore, in~\cite{KS08sono2}, this restoration has been confirmed explicitly 
by the Monte-Carlo simulation.  
\subsection{Simulation}
We have adopted the rational hybrid Monte-Carlo algorithm~\cite{CKS04}. 
We have use the code~\cite{remez} based on the Remez algorithm to 
find necessary coefficients in the simulation. 
In 2d ${\cal N}=(2,2)$ SYM, the complex phase of the fermion determinant 
is absent in the continuum limit and at discretized level only small phase appears 
as a lattice artifact. In this work we have ignored it. 
Fermi QCD/MDP~\cite{DiPierro:00} has been used to develop the simulation code. 

Because of the limitation of the resources, we have concentrated on 
the square torus, $L_x=L_y=L$. We took the number of sites and lattice spacings 
in two directions to be the same.  
For each set of parameters, we have collected 1000 $-$ 2000 samples of configurations.   
We have evaluated the error by using the Jack Knife method and it turned out 
the autocorrelations are sufficiently small. 


\end{document}